\documentclass[useAMS,usenatbib,]{mn2e}
\usepackage{graphicx}
\usepackage{color}
\usepackage[usenames,dvipsnames,svgnames,table]{xcolor}

\def\mpcnoh{\rm{Mpc}}

\def\lsim{\mathrel{\hbox{\rlap{\hbox{\lower4pt\hbox{$\sim$}}}\hbox{$<$}}}}
\def\gsim{\mathrel{\hbox{\rlap{\hbox{\lower4pt\hbox{$\sim$}}}\hbox{$>$}}}}
\def\kms {\rm{km~s^{-1}}}

\title{An improved method for the identification of galaxy systems: Measuring the gravitational redshift by Dark Matter Haloes}
\author[Mariano Javier de Le\'on Dom\'{\i}nguez Romero]{Mariano Javier de Le\'on Dom\'{\i}nguez Romero$^{1}$\thanks{E-mail:
mardom@oac.uncor.edu}, Diego Garc\'{\i}a Lambas$^{1}$ and Hern\'an Muriel$^{1}$\\
$^{1}$Instituto de Astronom\'ia Te\'orica y Experimental (IATE), Consejo de Investigaciones Cient\'ificas y T\'ecnicas de la Rep\'ublica Argentina (CONICET),\\
Observatorio Astron\'omico, Universidad Nacional de C\'ordoba, Laprida 854, X5000BGR, C\'ordoba, Argentina.}
\date{Accepted XXX Received XXX ; in original form \today}  
\pagerange{\pageref{firstpage}--\pageref{lastpage}} \pubyear{2012}
\begin{document}
\maketitle
\begin{abstract}
 We introduce  a new method for the identification of galaxy systems in redshift surveys based 
 on the halo model. This method is a modified version of the K-means identification algorithm developed by \cite{yang:04}.
 We have calibrated and tested our algorithms using mock catalogs generated using the Millennium simulations \citep{springel:05}
 and applied them to the NYU-DR7 galaxy catalog (based on the SDSS datasets).  Using this local sample of groups and clusters of galaxies we
 have measured the effect of gravitational redshift produced by their host dark matter haloes. Our results shows radial velocity 
 decrements consistent with general relativity predictions and previous measurements by \cite{whh:11} in cluster of galaxies. 
\end{abstract}
\begin{keywords}
galaxies: groups - galaxies: haloes - gravitation - cosmology: dark matter
\end{keywords}
\newpage
\section{INTRODUCTION} 
In a recent article \citep{whh:11} (WHH) analyze the pattern of spectroscopic redshifts for galaxies in 7800 clusters
 from the 7th data release of the Sloan Digital Survey (SDSS). Motion of a galaxy in a cluster generates a red or blueshift of its 
spectral lines; the equivalent velocities are typically on the order of 600 kms/s. WHH show that it is possible to disentangle 
the superimposed gravitational redshift predicted by the General relativity that corresponds to a radial velocity differences
 of the order 10 km/s. In their analysis the authors used a stacking data from several clusters and determined the shift of the 
redshift distribution's centroid with  growing radial coordinate.
The positions and redshifts of cluster galaxies are derived from a Gaussian Mixture Brightest Cluster Galaxy cluster catalogue \citep{hao:10}, 
one of the largest samples of galaxy clusters assembled on the basis of the SDSS-DR7. \\ 
In spite of the fact that massive clusters are expected to have the largest gravitational redshift effects, these systems are also affected
 by large galaxy peculiar velocities and substructure. Kim and Croft \citep{kc:04} suggested  that it is possible to overcome partially these
 difficulties by averaging over many clusters and groups of relatively low mass.\\
The existence of volume complete samples of galaxies with redshifts measured in the nearby universe as in the SDSS main galaxy sample \citep{abazajian:09};
 lead us to explore the gravitational redshift effects using systems of galaxies in the nearby universe.
Several groups have derived  catalogs of systems of galaxies applying  different techniques (Friends of friends FoF, Matched Filters etc.) to
 large data set such as the 2dF and the SDSS.\\
In order to make a meaningful comparison between observation and  theory, group definitions must have a reliable three dimensional counterpart.
 From the point  of view of  current theory of galaxy  formation,  the most direct route  for defining  galaxy groups is  dark matter halos. 
 More important for this work is the novel hybrid technique introduced by \cite{yang:04},  that starts with cluster candidates selected by
 a classical FoF  algorithm and a selected sample of bright isolated galaxies.
Other satellite galaxies are classified according a Maximum Likelihood criterion. Such method employs a model of the redshift space distribution 
of the galaxies (as function of system mass) as filter. A convolution of the model with the galaxy distribution allows to decide the membership of 
each galaxy to the proposed systems (provided by a suitable and calibrated FoF method) using an iterative procedure. 
The method relies in an assumed mass to light relation, which is a more reliable  method for compute the system mass that those based on the 
computation of the velocity dispersion and virial radii from galaxy positions.\\  
\cite{yang:04} method is a basic version of the well known K-means, a clustering algorithm where points are assigned to exactly one cluster and all
 points assigned to a cluster are equals in that system. Yang have carefully tested the performance of their group finder. The method was applied to
 the 2dFGRS  and compared with those extracted from detailed mock galaxy  redshift  surveys.  More recently \cite{yang:07} applied this method to the
 fourth release of the SDSS survey in order to study the dependence of color, stellar mass, star formation rate and morphology on halo mass.\\
In order to measure the gravitational redshift caused by dark matter haloes in systems of galaxies, we present in Section 2 modifications of the 
algorithm consisting in the introduction of a soft degree of assignment  of a galaxy to each cluster.  We have tested and calibrated
 our modified method in Section 3 with an extensive use of a galaxy mock catalog building up on the results of the Millenium simulations which 
mimics the  SDSS-DR7 galaxy catalog.
In section 4  we use a selected sample of clusters and groups of galaxies from the SDSS-NYU-DR7 datasets \citep{blanton:nyu} to 
compute the gravitational redshift produced by its dark matter haloes.  This is followed by a brief summary.
 We adopt for our study a flat concordance $\Lambda CDM$  cosmology \citep{komatsu:11} with $H_0= 72 \kms \mpcnoh, \Omega_M=0.26,
 \Omega_\Lambda=0.74$,  where appropiate we define $h=H/100/\kms/\mpcnoh$.\\ 
\section{IDENTIFICATION OF SYSTEMS OF GALAXIES:}
The aim of this section is to develop a group finder that assigns galaxies in a common halo to a single group. 
We will follow the methodology of \cite{yang:04} which has advantages over other group identification algorithms, and 
 introduce some improvements related to the K-means implementation as well as the initial seeds for the model.
%
%
%
%
This method relies on the introduction of a background level B, that defines a threshold density contrast in redshift space. Ideally this
 value should correspond roughly to the redshift-space contrast at the edge of a halo. Taking into account theoretical considerations, Yang
 et al obtain $B \sim 10$, in excellent agreement with the value  obtained by maximizing the completeness and minimizing the contamination.\\
 Nevertheless, we first notice that it is not necessary to introduce such a high threshold galaxy density contrast.
 Rather we allow a lower background level value so that the likelihood of a galaxy in a given halo is greater than that of the
 main or central galaxy of another nearby haloes. Such value was chosen using the simulated catalogs described below.
 We can therefore use the likelihood value as main/satellite classification criteria by classifying a 
galaxy as satellite if the likelihood of pertenence to another halo is over the background level.
We argue that this methodology could be used as a classification process. Rather than starting from prefixed seeds halos 
(such as in the  work of Yang et al.), each galaxy in the catalog is allowed to start as central galaxy in the halo
 where it resides. As the iterative method proceed, many galaxies subsequently will be classified as satellites.\\
No further criteria for the seeds selection are needed with this improvement. This avoid the problems introduced
by the  pre-selection of seeds  using a FoF algorithm ( discussed by \cite{berlind:06}).
%
%
\subsection{From hard to soft assignments:}
Yang's cluster assignment method can be considered as an example of a competitive learning algorithm: the well known K-means.
The K-means is an algorithm for assigning N data points (i.e. galaxies) into K clusters (i.e.  groups of galaxies). 
 Each $k$ cluster is parameterized by a vector ${\vec \Omega}^{k}$ (which is called the cluster mean) with elements position of the 
center and size of the group of galaxies.  The algorithm works by allowing the K-clusters to compete each other for the right to own the data points.
%
This procedure owes its advantage in its simplicity, and in its local-minimum convergence properties. 
However it as a main shortcoming, is a "hard" rather than "soft" algorithm: points assigned to a cluster are equals
 in  that cluster. Points located near the border between two or more clusters should also play a role in determining
 the locations of all the clusters that they could plausibly be assigned to. Nevertheless in the K-means algorithm, each borderline
point is dumped in one cluster, with the same vote than all the others points in that cluster, and no vote 
in any other cluster.  The previous criticisms of K-means motivate the "Soft K-means algorithm" used in this work (see \cite{MacKayBook}). \\
A fundamental hypothesis for this clustering problem is the underlying idea that the maximum likelihood method, which identifies the setting 
of the parameter vector $\vec{\Omega}$ that maximizes the likelihood $P({\rm{Data}}| \vec{\Omega}, {\rm{Model}})$ provides a good fit to the data.
 Briefly, the algorithm is set in three steps: Initialization, Assignment and Update.\\
{\bf Step 1: Initialization} In order to start the K-means, the mean values {${\vec \Omega}^{(k)}$} must be initialized
 in some way, we start using all the galaxies in the catalog. 
%
A representation of this assignment of galaxies to groups is given by the so called "responsibilities", which are indicator 
variables $r_{k}^{(n)}$ that a given galaxy $n$ belongs partially to the $k$ group of galaxies.  We firstly take all 
galaxies  as `central', and so are considered as the center of a potential dark matter halo.
 Using the total group luminosity $L_{\rm  group}$ and a  model  for  the  group mass-to-light ratio  we can estimate 
 the halo mass associated i.e.  the halo radius $r_{\rm 200}$, the virial radius $r_{\rm  vir}$, the  virial 
velocity $V_{\rm  vir} = (G  M / r_{\rm  vir})^{1/2}$ and the line-of-sight velocity dispersion  of the galaxies within the 
dark matter halo  $\sigma = V_{\rm vir}/\sqrt{2}$.
 The total luminosity of a selected potential group is estimated using (at this stage is simply the luminosity of the galaxy).
%
$L_{\rm group} = \sum_i {r_{k}^{(i)} L_i},$
%
where $L_i$ is  the luminosity of each galaxy in  the group, $r_{k}^{(i)}$ is the k group responsibility for each (i) galaxy (=1 initially). 

{\bf{Step 2: Assignment step}} Once we have a group centre, and a tentative estimate of the group size,  mass, and velocity dispersion,
 we  can assign satellite galaxies to this  group according to the properties of the associated halos. 
If  we assume that  the phase-space  distribution of  galaxies follows that  of the  dark matter  particles, the  number density  contrast of
galaxies in redshift space $P({\rm{Data}}|\vec{\Omega^{k}}, {\rm{Model}})$ around the k group centre at redshift $z_{\rm group}$ is computed 
in a similar way than in Yang's technique.
\begin{equation}
P({\rm{Data}}|\vec{\Omega^{k}}, {\rm{Model}})= P_M(R,\Delta z) = {H_0\over c}
{\Sigma(R)\over {\bar \rho}} p(\Delta z) \,,
\end{equation}
Here $\Delta z  = z - z_{\rm group}$ and  $\Sigma(R)$ is the projected surface density of a (spherical) NFW halo.  Hereafter $R$ is the projected
 distance at the redshift of the group. 
The  function $p(\Delta  z){\rm d}\Delta  z $  describes  the redshift distribution of galaxies within the halo for which standard Gaussian
shape is adopted, with $\sigma$ as the rest-frame galaxy velocity dispersion.
Thus  defined,  $P_M(R,\Delta  z)$  is the  three-dimensional  density contrast  in redshift  space.  In  order  to decide  whether a given
 galaxy should be assigned as a primary or as a satellite into a particular group, for each galaxy  we loop  over all groups,  and  compute  the corresponding
 distance $(R,\Delta z)$  in phase space between galaxy and group centre. 
Firstly we classify the galaxies as primaries or satellites according to the value of $P({\rm{Data}}|\vec{\Omega}, {\rm{Model}})$ starting from the most massive 
systems towards the smaller ones, as was described in subsection 2.1. Instead of assigning each galaxy (n) to the group for
 which $P_M(R,\Delta z)$ has the highest value, we proceed to evaluate the responsibilities for each group (k) as: 
\begin{equation} 
r_{{k}}^{(n)}= \frac{ \pi_{(k)} P({\rm{Data}}| \vec{\Omega^{k}}, {\rm{Model}})}{ \sum_{\acute{k}} \pi_{(\acute{k})} 
P({\rm{Data}}| {\vec{\Omega^{\acute{k}}}}, {\rm{Model}}) }
\end{equation} 
{\bf{Step 3: Update step}} 
In the update step, the following model parameters  ($\vec{\Omega}^{k}$): center angular position and group redshift,
 are adjusted to match the sample mean of the data points ${\vec x}^{(n)}$ that they are responsible for      
${\vec{\Omega}}^{k}=\frac{\sum_{n} r_{k}^{(n)} {\vec x}^{(n)}}{R^{(k)}}$, 
where $R^{(k)}=\sum_{n} r_{k}^{(n)}$ is the total responsibility of mean $k$.
%
Now the algorithm also includes groups weighting parameters $\pi_{(k)}=\frac{R^{(k)}}{\sum_{k} R^{(k)}}$ which also update themselves, 
allowing accurate modeling of data from groups of unequal weights.
%
The total responsibility of a cluster becomes a better measure of the real occupation of the halo, and being a continuous parameter it is more
adequate for modeling and interpretation.  Since some of the groups parameters (mass, velocity dispersion, virial radii) are derived from the total group luminosity,
we could made use of the responsibilities in order to obtain a "soft" weighted luminosity, which should be more representative of the real value.
This procedure has the advantage that the inferred luminosities (i.e. the masses) of the groups are almost independent of the possible contamination by interlopers. 
It should be recalled that the memberships of the selected groups  are remarkably insensitive to the adopted mass-luminosity model, as was stated by \citep{yang:04}.
%
Next, we iterate between Step 2 and 3 until  there are  no further changes in  the memberships. Finally we assign membership for each galaxy to
 the group which have the largest responsability (in case of a tie, we assign to the most massive group).\\ 
Note  that, unlike with the usual FOF method, this group finder also identifies groups with only one member. Since we start using all the galaxies
 in the catalog, we could potentially assign each galaxy to their corresponding dark matter halo, since we include all possible groups seeds.
\section{Application to Simulated and SDSS Catalogs}
Numerous mock catalogs have been produced from full-blown semi-analytic model of galaxy formation \citep{wang:06}.
The \emph{Millennium Simulation} \citep{springel:05} used in this work, is one of the largest simulation of cosmic structure growth carried out so far.
%
In what follows,  we will use the semi-analytic galaxy catalogues at redshift zero constructed from the Millennium simulation by
\cite{croton:05} (http://www.mpa-garching.mpg.de/galform/agnpaper/),  who considered a detailed model for cooling, star
 formation, supernova feedback, galaxy mergers and metal enrichment as well as  a simple treatment of heating by a central AGN.
These authors find that several  observed properties of the galaxy population at $z\!=\!0$ are suitably reproduced indicating that 
galaxy characteristics and stellar ages are in much better agreement with observation than for models without AGN feedback.
We build up mock catalogs of the SDSS-DR7 (spectroscopic) survey in order to calibrate our system
 identification method as well as comparing observed galaxy properties of models and observations.
Since the median redshift of galaxies in SDSS is $\sim 0.1$, the $r$-band absolute magnitudes $M_r$ of each model galaxy
are K-corrected to $z=0.1$ ($M^{0.1}_r$) using the $K-$correction code ({\tt kcorrect v3\_1b}) of \citet{blanton:07} and
 the luminosity evolution model of Blanton et al. (2003). 
Galaxy redshifts are assigned by placing the observer at the corner of the simulation box and are determined by the comoving distance 
to the observer plus the galaxy peculiar velocity contribution.  The corrected $r$- band magnitudes are given  by:
%
${M^{0.1}_r}= {{-2.5}{\times}{\log}{L}+{K_{correction}}+{E_{correction}}-{5}{\log}{h}}$.\\
%
The survey geometry and the radial selection function were properly taken into account in the construction of the mock catalogue where
 we record galaxy redshifts, angular coordinates,  3D positions, apparent and absolute u,g,r,i and z band SDSS magnitudes, as well as an index 
to identify the galaxy assignment to a dark matter halo.
In order to model the mass-weighted luminosity relation we select a common luminosity scale $L_{19}$, defined as the luminosity of all group members brighter than
$M_{r}^{0.1}=-19+5*log(h)$. To calibrate this relation for the weighted luminosities we select from the mock SDSS catalog all groups with $z < 0.068$, corresponding to
the volume limited magnitude given for a catalog with apparent magnitude limit of $17.5$.
Using the actual group centers and masses we compute the responsibilities of each group over all the galaxies in the catalogue.  The use of responsibility
 weighted luminosities provides a slightly improved (with lower dispersion) mass-luminosity relation, compared to \cite{yang:04}.
Since  the total group luminosity  is dominated by the few brightest galaxies, the estimated mass is much less sensitive to the
 absence of faint group members (typical in flux-limited catalogs) and the effects of interlopers.
This methodology allows to avoid one of the problem of the maximum likelihood, the well known phenomenon of overfitting.  
For all groups selected below the redshift limit $z < 0.068$ we compute $L_{19}$ directly from the selected members with 
$M_{r}^{0.1}-5*log(h) \le -19.0$. For groups at higher redhifts, we compute $L_{\rm group}$ and use the average relation 
between $L_{19}$ and $L_{\rm group}$ to estimate the former.
\subsection{Performance of the method}
\begin{figure}
\centering
\includegraphics[width=8.5cm,height=5.cm]{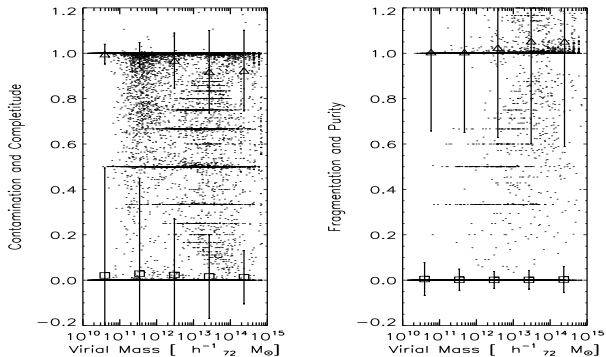}
\caption{Left panel: completeness (open triangles) and contamination (open squares), Right panel: fragmentation (open squares)
and purity (open triangles) of groups (averages) as a function of halo mass for systems identified in the mock catalogue based on
 the Millennium simulations. Errorbars correspond to 1-$\sigma$ variance.
}
\label{fig:1}
\end{figure}

A {\it real} group is defined as  the set of galaxies that reside in  the same halo of the mock catalog. In order to  quantify the group finder
performance we introduce the completeness, defined as the ratio between the number of  true members selected by the group finder
and the  total number  of true group  members, and  the contamination, defined as the  ratio between the number of false members (interlopers) selected  by the 
 group finder and  the total  number of real members.  We also measure the fragmentation,  defined as the mean number of extra galaxy groups per dark 
matter halo having mass at least 0.1 times that of their true associated galaxy group and compute the purity of the systems defined as the
 ratio between the true members and the group members.\\
Figures \ref{fig:1} gives an overview of the completeness, contamination, fragmentation and purity for our group finder. Dots correspond to individual groups
(halos), while large triangles indicate the average for all the systems included inside the window function without restrictions in the
absolute magnitude limits.  As can be observed, for high mass/population systems it is difficult to recover all the members in the outer regions. 
Nevertheless, these objects are preferentially infalling galaxies that are difficult to recover by any method.\\
Fragmentation of groups should be considered if these subgroups comprise a considerable mass
fraction. If this fraction is $\sim 10\%$ the effect is negligible for high mass systems but affects intermediate mass/occupation systems. 
%
With the modifications introduced, the improved algorithm is significantly more efficient than the popular FoF identification algorithms 
or its variants in terms of high completeness and purity, low contamination and fragmentation as is shown in Figure  \ref{fig:1}.
Notice that our test are based on a mock sample from a full N-body SAM based model, which presumably results in a more realistic scenario than a CLF
 based one as is the 2dF mock catalog used by \cite{yang:07}.
Three important output parameters of the method should be stressed: each group responsibility for a galaxy provides an effective medium for interlopers 
classification (as an example a galaxy with similar responsibilities by two different and nearby haloes);  the total responsibility of each halo, which 
could serve as a new halo occupation measure (introducing some limit on magnitude); and the total weighted luminosity as an important intermediate step to the 
computation of the total dark matter halo mass. 
\subsection{System Identification in the SDSS-NYU-DR7}
Using a dedicated 2.5 meter telescope, SDSS provide the largest and most complete photometric and spectroscopic galaxy survey available.
 We have used the large-scale structure sample \texttt{sample14} from the NYU Value Added Galaxy Catalog (NYU-VAGC; 
\cite{blanton:nyu}) DR7 version as our primary galaxy sample, a complete sample of galaxies with reddening corrected r magnitudes brighter than 17.5.
We take into account fiber collisions by giving each collided galaxy their photometric redshift estimate.
In order to limit the effects of incompleteness on our group identification procedure, we
restrict our sample to regions of the sky where the completeness (ratio of obtained redshifts to spectroscopic targets) is greater than $90\%$.  
Galaxy magnitudes are corrected for Galactic extinction using the dust maps of \citep{schlegel:98}, absolute magnitudes are k-corrected
\citep{blanton:07} and corrected for passive evolution \citep{blanton:03} to rest-frame magnitudes at redshift $z=0.1$. 
We use the fit $Q(z)=1.6(z-0.1)$, which takes into account the effects of the luminosity evolution. 
We  select all galaxies with extinction corrected apparent magnitude  brighter than  $r = 17.5$, redshifts in the
range  $0.015 \leq z \leq 0.15$, and with redshift measurement completeness $\ge  0.9$. We use a lower redshift limit $z>0.015$ to alleviate
some of the problems asociated with the accuracy of the photometry of some nearby extended galaxies.  
For a uniform magnitude limit, our selection function  $\bar{n({\bf{r}})}$, can be separated 
into the product of an angular and a radial part: $\bar{n({\bf{r}})}=\bar{n({\bf{\hat{r}}})}\bar{n(r)}$,
where ${\bf{R}}\equiv r {\bf{\hat{r}}}$ and $ {\bf{\hat{r}}}$ is a unit vector. The angular part may take any value between 0 and 1, 
and provides the completeness as a function of position, i.e., the fraction of all survey selected galaxies for which reliable redshifts are given.
In order to compute the selection function, we use Mangle, {\bf{http: //casa.colorado.edu/~ajsh/mangle}}.  Such codes \citep{hamilton:04}, are
 designed to deal accurately and efficiently with complex angular mask.  We apply the procedure suggested by \cite{berlind:06}) to the groups identified
 by our algorithm on the NYU-SDSS-DR7 galaxy sample using the mask in order to avoid completeness issues due to the angular footprint. 
\section{A Measure of Gravitational Redshifts}
\begin{figure}
\begin{center}$
\begin{array}{cc}
\includegraphics[width=1.5in]{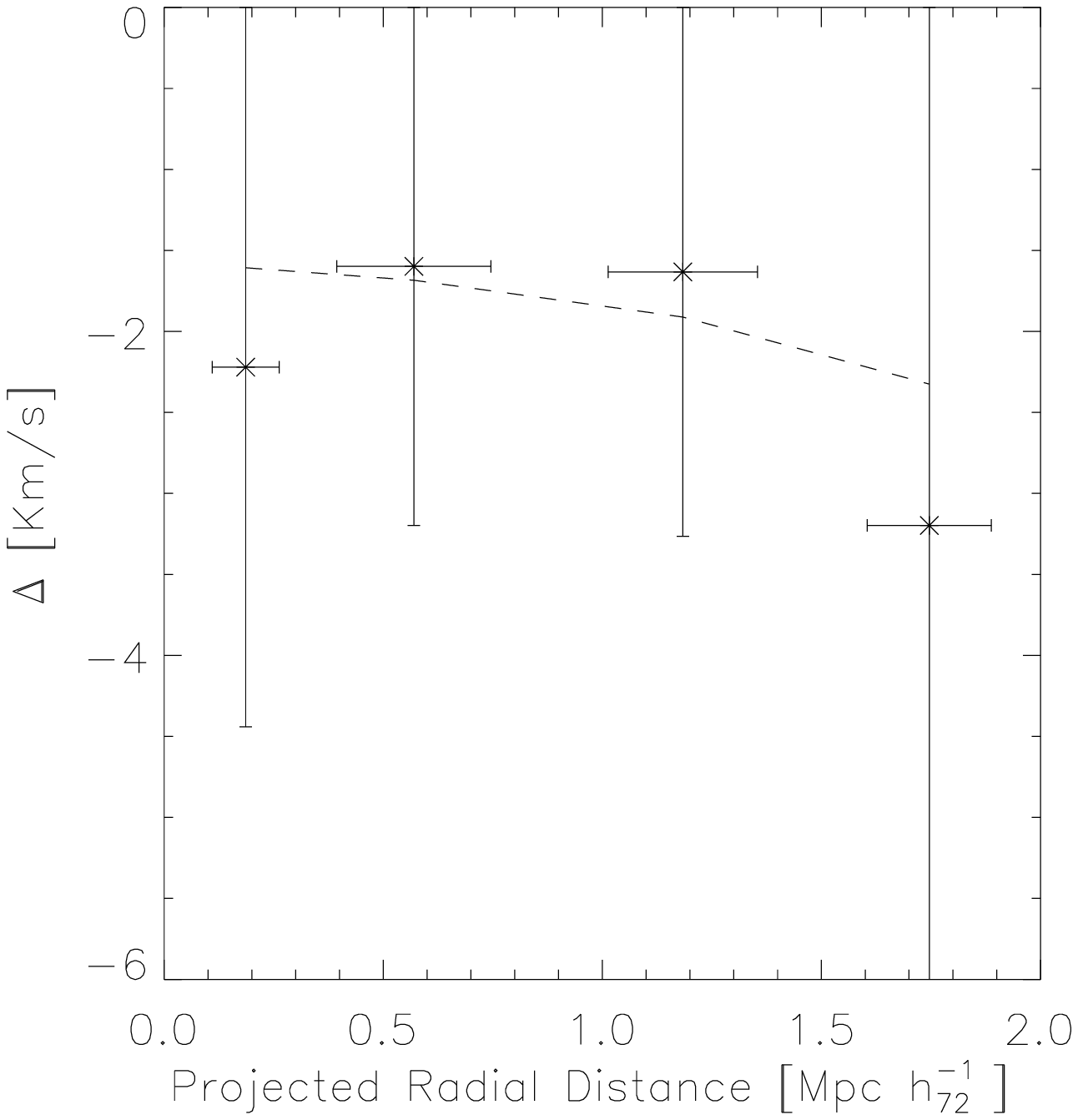} & 
\includegraphics[width=1.5in]{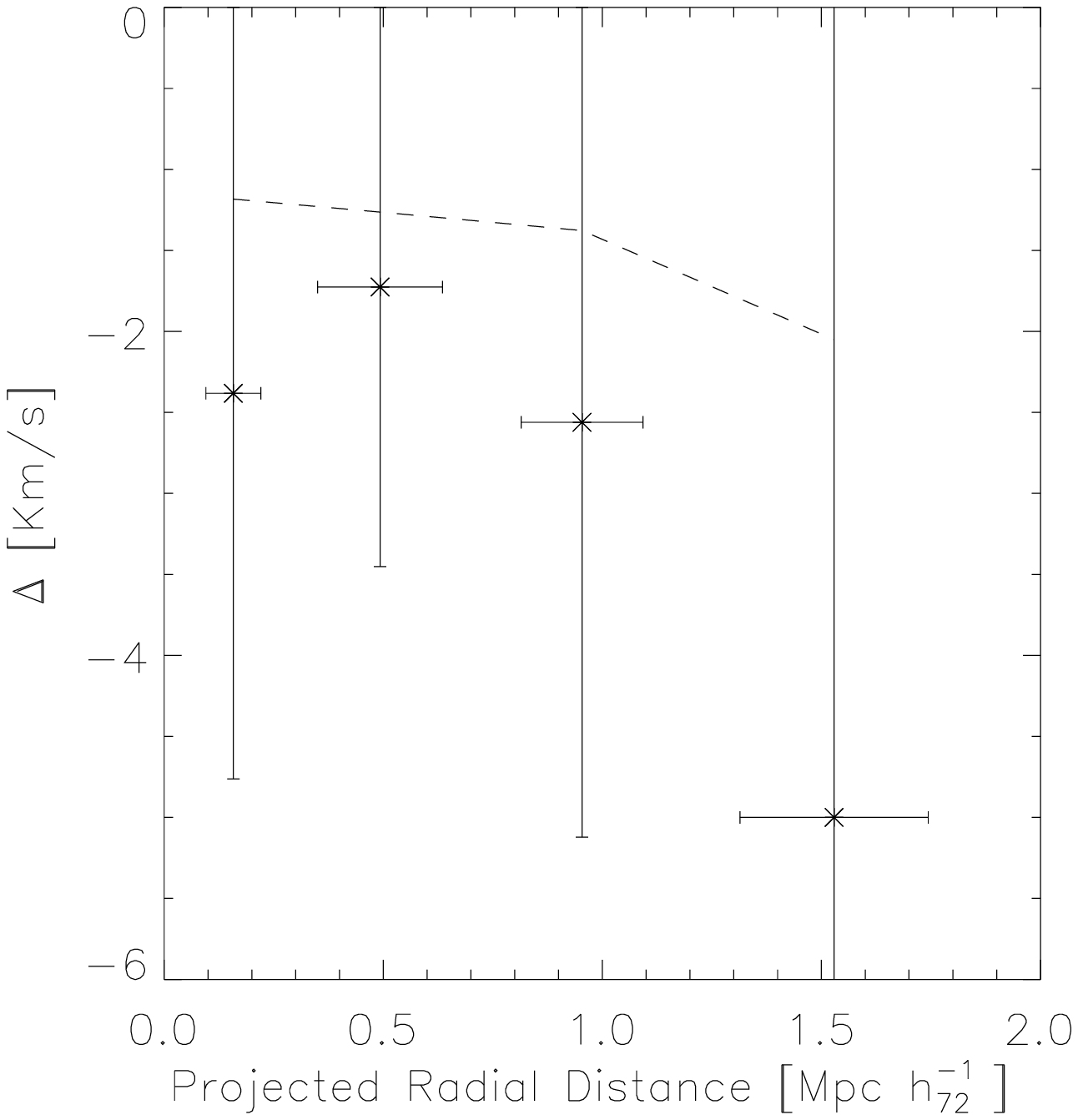} \\ 
\includegraphics[width=1.5in]{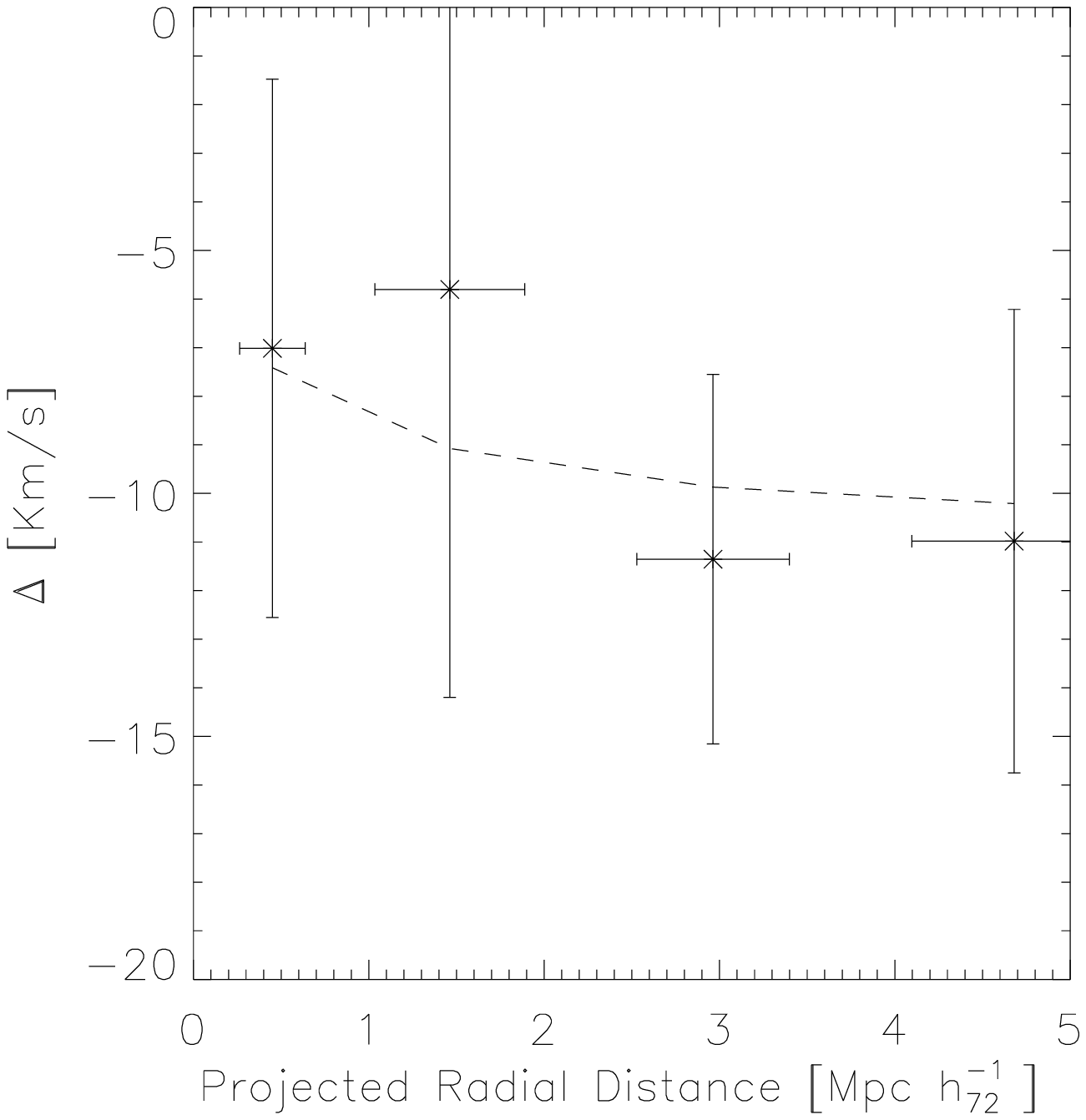} &
\includegraphics[width=1.5in]{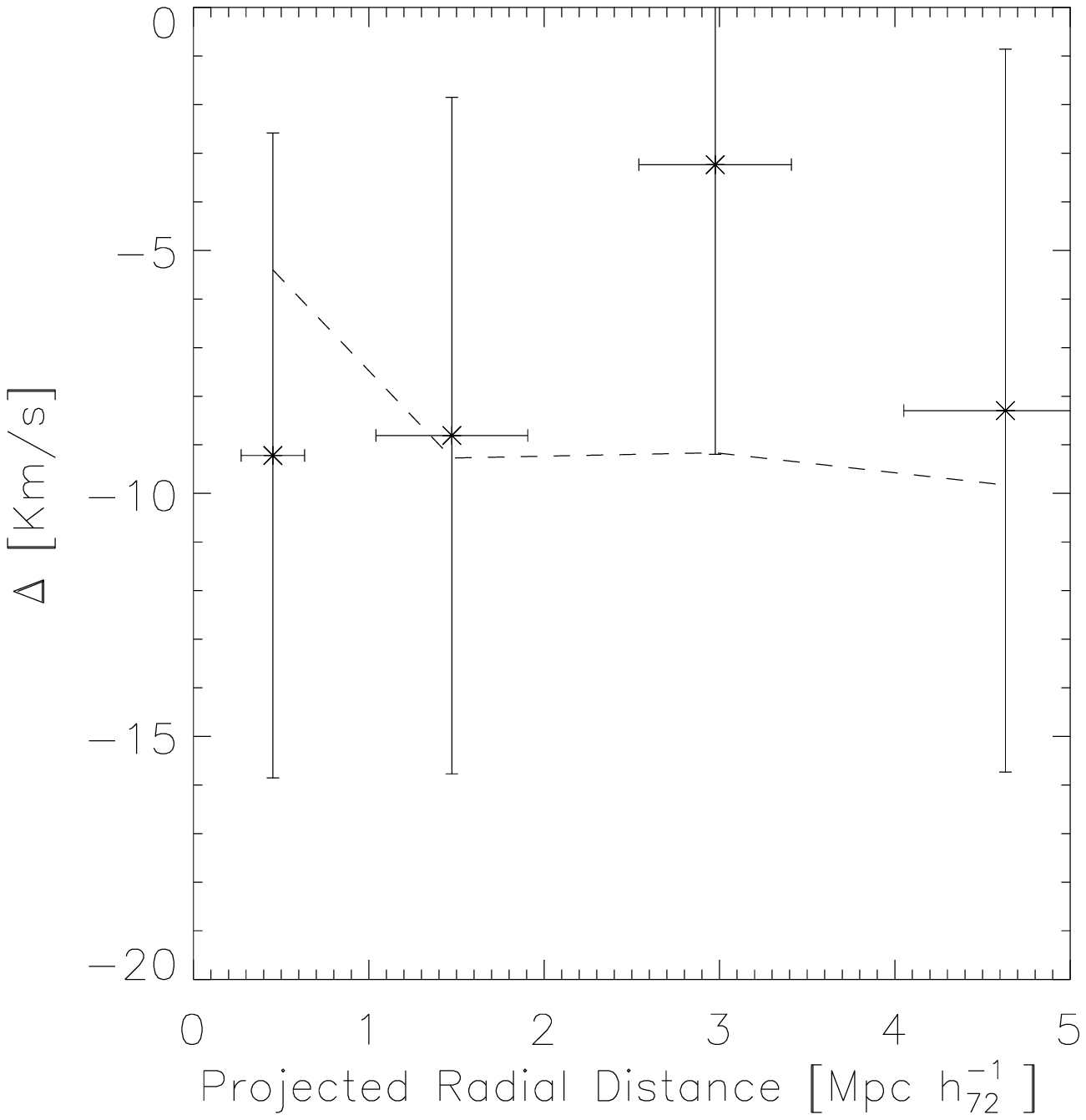}
\end{array}$
\end{center}
\caption{Observed gravitational redshift (points with 1-$\sigma$ errorbars)in terms of the mean velocity $\Delta$, as a function of the projected
clustercentric distance for Millennium simulated systems in the mock catalog (left panels) and SDSS-NYYU-DR7 systems of galaxies (right panel). 
Upper and lower panels correspond to groups and clusters of galaxies respectively. Dashed lines show the General Relativity 
theoretical predictions.
} 
\label{fig:2}
\end{figure}

Using the systems of galaxies identified in the SDSS-NYU-DR7 galaxy redshift survey we measured the gravitational redshift 
due to the dark matter gravitational profile. A standard stacking technique is used in order to disentangle the kinematic
Doppler effect from the gravitational redshift, given that the latter shifts the centroid of the observed velocity distribution.
Coordinates and redshifts of cluster centres need no corrections since our identification methodology provides reliable
values. We search for all galaxies within a window in phase space of $3$ virial radii and $3 \sigma$ in velocity around the cluster centres. 
 As the final step we combine redshift data of all clusters into one. Since we are interested in the detection of
the GR effect in intermediate mass systems, we have considered two samples of systems of galaxies: low mass ($10^{13} M_{\odot}$ to $10^{14} M_{\odot}$, 
 groups of galaxies) and high mass ($\> 10^{14} M_{\odot}$, comparable to the cluster sample analyzed by WHH). 
Our final sample comprises systems of galaxies in the range $0.015 \leq z \leq 0.15$ with a mean redshift of $0.1$ hosting on average $28$ galaxies with 
spectroscopic redshift measurements for clusters and $14$ for groups. Those {\rm{groups}} with less than $4$ redshift determinations were not included 
in the sample.
Since the presence of interlopers in our sample is negligibly small, it is possible to measure directly the mean value of 
the velocity distribution of cluster galaxies, $\Delta$. Assuming spherical symmetry and no strong inhomogeneities, the gravitational 
redshift profile of a system can be calculated using the following formula \citep{cappi:95}
\begin{equation}\label{vel-shift-proj}
\Delta_{\rm s} (R)=\frac{2}{c\Sigma(R)}\int_{R}^{\infty}[\Phi(r)-\Phi(0)]\frac{\rho(r)r\textrm{d}r}
{\sqrt{r^2-R^2}},
\end{equation}
where $R$ is the projected cluster-centric distance, $\Phi(r)$ is the gravitational potential, 
$\rho(r)$ and $\Sigma(R)$ are the 3D and surface (2D) density profiles of galaxies. In order to estimate 
this effect for the data combined from a cluster sample, it is needed to convolve this expression with 
the distribution of cluster masses in the sample. In order to test the possibility of determining this effect
in our sample we introduce the corresponding blueshift in the mock catalog of galaxies in their parent haloes. 
Given that the dark matter haloes follow an NFW density profile \citep{navarro:96}, the gravitational potential
results:
\begin{equation}\label{psiNFW}
\Phi(r)=-(GM_{\rm v}/r_{\rm v})^{1/2}g(c_{\rm v})^{1/2}\frac{\ln(1+r/r_{\rm v})}{r/r_{\rm v}}
\end{equation}
and $g(c_{\rm v})=1/[\ln(1+c_{\rm v})-c_{\rm v}/(1+c_{\rm v})]$ with $c_{v}$ the concentration depending on the halo mass.
 Using formulas 3 and 4 we introduce the corresponding gravitational redshift effect by adding $\Delta$ to the
 line of sigth velocities of the member galaxies (Hubble flow plus peculiar velocities) that populated each dark matter halo in the Millennium simulation.
In the left panel of figure \ref{fig:2} we show $\Delta$ as a function of the projected radial distance and its comparison 
to the imposed effect in the systems identified in our mock catalogs. 
This should be compared with the measurement showed in the right panel for the SDSS-NYU-DR7 systems.  Upper/Lower panels correspond
to low/high mass systems. Although the predicted signal is significantly smaller 
for groups than for clusters of galaxies (showed in dashed lines), the quality and size of the sample of low mass systems made 
possible to clearly detect the gravitational redshift effect predicted by the General Relativity. We notice that the high mass results are in agreement  
with \citep{whh:11}.
\section{CONCLUSIONS}
\begin{itemize}
\item An efficient halo finding algorithm has been developed and tested using mock catalogs derived from numerical simulations with 
semianalytic galaxy formation methods.  The performance of the method is solid and efficiently provides complete, almost bias free 
galaxy system catalogs based on galaxy redshift surveys.
\item The new method was used to compile samples of groups and clusters of galaxies in the SDSS-NYU-DR7. The clusters of galaxies
 sample was used to reproduce the results found by WHH in clusters of galaxies. Following the suggestion by \citep{kc:04}, we are able 
to detect the gravitational redshift in the sample of groups of galaxies. 
\item Our measurements of the gravitational redshift effect by dark matter haloes show a good agreement with the predictions of General Relativity.
\end{itemize}
%
\bibliographystyle{mn2e}%
\bibliography{lettergr.bib}
%
\section*{Acknowledgments}
We thank the referee for their comments and helpful suggestions for manuscript changes. These have improved
 both the content and clarity of the manuscript.
 This work has been partially supported by Consejo de Investigaciones Cient\'{\i}ficas y T\'ecnicas de la Rep\'ublica Argentina (CONICET) and the SeCyT-UNC.
 This research has made use of NASA's Astrophysics Data System.
 The Millennium Run simulation used in this paper was carried out by the Virgo Supercomputing Consortium at the Computing Centre of the Max-Planck Society in Garching. 
The semi-analytic galaxy catalogue is publicly available at http://www.mpa-garching.mpg.de/galform/agnpaper".
 Funding for the creation and distribution of the SDSS Archive has been provided by the Alfred P. Sloan Foundation, the Participating Institutions, the  NASA, the NSF, the U.S. DoE, the JM, and the MPS.
 The SDSS Web site is http://www.sdss.org/.
\end{document}